 \documentclass[preprint]{aastex}

\slugcomment{To appear in the Astrophysical Journal}

\shorttitle{CH$_{\rm 3}$OH in IRAM 04191+1522} 
\shortauthors{Takakuwa et al.}

\begin{document}


\title{Interaction between the Outflow and the Core in IRAM 04191+1522}

\author{Shigehisa Takakuwa\altaffilmark{1,2}, Nagayoshi Ohashi, \& 
Naomi Hirano}
\affil{Academia Sinica Institute of Astronomy and Astrophysics,\\ 
P.O. Box 23-141, Taipei 106, Taiwan} 
\altaffiltext{1}{e-mail: stakakuwa@sma.hawaii.edu}
\altaffiltext{2}{Present address: Harvard-Smithsonian Center for 
Astrophysics, 82 Pu'uhonu Place, Suite 210, Hilo, HI 96720, U.S.A.}

\begin{abstract}
We have carried out mapping observations of the 
molecular core associated with the young Class 0 protostar, IRAM 
04191+1522, in the CH$_{\rm 3}$OH 
($J_{\rm K}$=2$_{\rm K}$--1$_{\rm K}$) and C$^{\rm 34}$S ($J$=2--1) lines 
using the 45 m telescope at Nobeyama Radio Observatory. 
Our observations have revealed that 
there is a condensation associated with the protostar, elongated 
in the east-west direction mostly perpendicular to the axis of the 
associated CO outflow. Its size and mass are estimated to be 
0.07 pc $\times$ 0.04 pc 
and 2.3 M$_{\odot}$, respectively, from the CH$_{\rm 3}$OH data. 
In addition to the elongated envelope, 
two compact ($\sim$ 0.03 pc) condensations were found 
in the CH$_{\rm 3}$OH line 
at the southern edge of the elongated envelope, where
the blueshifted CO outflow emerging from the protostar is located.
In contrast to the elongated envelope, 
those compact CH$_{\rm 3}$OH condensations show much larger 
line width (up to 2.0 km s$^{\rm -1}$) with centroid velocities
blueshifted by $\sim$ 0.8 km s$^{\rm -1}$. 
The compact condensations have momenta ($\sim$ 0.06 M$_{\odot}$ 
km s$^{\rm -1}$) comparable to that of 
the blueshifted molecular outflow. In addition, they are gravitationally
unbound, and most probably will dissipate eventually. 
These results suggest that the compact condensations are probably
formed in the course of interaction between the outflow and
the ambient gas surrounding the protostar, and that 
such interaction may cause dissipation of a part of the ambient gas.
No drastic, localized enhancement of the CH$_{\rm 3}$OH 
abundance is, however, observed toward the compact condensations,
implying that there seems to be no significant shock heating at the
compact condensations in spite of the interaction with the outflow.
This may be because the CO outflow velocity ($<$ 10 km s$^{\rm -1}$) 
is too low to cause 
effective heating to release CH$_{\rm 3}$OH on dust grains into gas phase.
\end{abstract}


\section{Introduction}

It is widely accepted that molecular outflows take place during 
low-mass protostellar formation. The origin of molecular outflows is still 
debatable, whereas recent observational and theoretical studies have 
suggested that high-velocity neutral jets from the vicinity of
protostars drive  
molecular outflows by entraining the material along the 
jet path \cite{rag93,dut97,gue98,ha99a,ha99b}. Those jets could cause 
shocks and alter the physical and chemical conditions of the entrained 
material drastically. In fact, there is observational evidence of such 
shocks from some Class 0 sources, 
such as L1157 
\cite{mik92,ume92,bac95,ave96,bac97,gue98,ume99,bac01}, NGC 1333 
IRAS 2 \cite{san94,bac98}, IRAS 4 \cite{bla95}, IRAS 16293-2422 
\cite{bla94,cec00,hir01}, L1448 \cite{bac91,dut97}, HH25MMS \cite{gib98}, 
and BHR71 \cite{bou97,gar98}. These sources show that, in the 
positions where jets hit and entrain the ambient dense gas, molecular 
lines such as SiO, SO, CS, and CH$_{\rm 3}$OH are much more intense 
than those in the positions of the protostars. These results have 
been interpreted as the case where molecules on dust grains evaporate 
into gas phase through the C-type shock heating or sputtering due to the 
interaction with jets, resulting in enormous abundance enhancement of 
these molecules \cite{ka96a,ka96b,hol97,pin97,sch97,ber98,buc02}. 

However, studies of the interaction between outflows and parent cloud 
cores have been so far limited to some representative sources with 
strong shocks like those mentioned above. 
Hence, we do not know how ubiquitous 
and how important the interaction between outflows and cores is in the 
process of low-mass star formation. For example, there is no thorough 
observational studies on this issue in the Taurus molecular cloud 
complex, even though it is one of the nearest (140 pc ; Elias 1978) 
and the most 
representative low-mass star-forming regions. 

IRAM 04191+1522 (hereafter referred to as IRAM 04191) is a Class 0 
protostar, recently discovered by Andr\'e et al. (1999a) in the southern 
part of the Taurus molecular cloud. This source, with a very low 
bolometric temperature (T$_{bol}$) of $\sim$ 18 K, is located at 
$\sim$ 1$\arcmin$ southwest of the Class I infrared source IRAS 
04191+1523 (not to be confused with IRAM 04191), and is considered 
to be one of the youngest protostars 
among the known Class 0 objects \cite{an99a,an99b}. 
IRAM 04191 is associated with a well-collimated molecular outflow 
that extends from northeast (red) to southwest (blue) \cite{an99a,lee02}. 
Since such an extremely young protostar 
should be surrounded by a plenty of material which is less disrupted by 
the associated outflow, we may expect a more clear case of the 
interaction between the outflow and the surrounding material. 
In fact, Belloche et al. (2002) have recently found evidence of such 
an interaction toward the blue lobe of the molecular outflow in 
the CS(2--1) line. 
It is therefore quite interesting to study the 
interaction between the outflow and the core toward this very young 
protostar in detail. 

We adopted a band of CH$_{\rm 3}$OH ($J_{\rm K}$=2$_{\rm K}$--1$_{\rm K}$) 
lines \cite{fri88} as a probe to study interaction between the outflow 
and the parent cloud core associated with IRAM 04191. As mentioned 
above, the CH$_{\rm 3}$OH line can be a good tracer of shock 
interaction, since CH$_{\rm 3}$OH molecules on dust grains evaporate 
into gas phase by mean of shock heating and the abundance is greatly 
enhanced toward those interaction regions 
\cite{san94,bac95,bla95,ave96,bac97,bac98,bac01}. In addition to the 
CH$_{\rm 3}$OH lines, the C$^{\rm 34}$S ($J$=2--1) line, the enhancement 
of which is also observed in some sources with outflow-core interaction 
\cite{san94,bla95,gar98,bac01}, 
was observed simultaneously. Belloche et al. (2002) also
observed C$^{\rm 34}$S ($J$=2--1) as well as the main isotope in IRAM04191.

\section{Observations}

We carried out mapping observations of the CH$_{\rm 3}$OH ($J_{\rm 
K}$=2$_{\rm K}$--1$_{\rm K}$) and C$^{\rm 34}$S ($J$=2--1) lines 
with the 45 m telescope at Nobeyama Radio 
Observatory\footnote{Nobeyama Radio Observatory (NRO) is a branch of 
the National Astronomical Observatory, an inter-university research 
institute operated by the Ministry of Education, Culture, Sports, 
Science and Technology of Japan.} from the 8th to the 11th of January, 2000. 
The mapping observations have been performed at a grid spacing of 
20$\arcsec$ and a 
position angle of 30$^{\circ}$, to align the mapping grid to the axis of 
the outflow. The observed molecular lines are summarized in Table 1. 
We used two SIS receivers with 500 MHz bandwidth 
by dividing the beam into two linear
polarized waves through a wire grid, each of which 
covered the frequency range of the molecular lines listed in Table 1, 
and co-added the data to improve the signal to noise ratio. 
The beam squint between the two receivers 
was confirmed to be less than 1$\arcsec$. Typical single sideband 
(SSB) system noise temperature was $\sim$ 300 K for both receivers. 
The beam size and the main beam efficiency of the telescope was 
17$\arcsec$ and 0.5, respectively, at 96 GHz. The telescope pointing 
was checked every 60 - 90 minutes by observing SiO maser emission 
from NML-Tau at 43 GHz, and was confirmed to be better than 
5$\arcsec$. The observational mode was position-switching. 

We used a bank of eight high-resolution acousto-optical spectrometers 
(AOS) with a bandwidth of 40 MHz and a frequency resolution of 37 kHz. 
The frequency resolution corresponds to $\sim$ 0.11 km s$^{\rm -1}$ 
for the observed lines. Typical rms noise per channel was $\sim$ 0.06 K 
in units of T$_{\rm A}$$^{\rm *}$ (hereafter we present line 
intensities in units of T$_{\rm A}$$^{\rm *}$).

In order to calibrate the intensity variation caused by different receiver 
gains and/or sky conditions, we observed the central position of IRAM 
04191 ($\alpha$$_{2000}$ = 04$^{\rm h}$ 21$^{\rm m}$ 56$^{\rm s}$.9, 
$\delta$$_{2000}$ = 15$^{\circ}$ 29$\arcmin$ 46$\arcsec$.4) in every 
scan. We confirmed that the intensity variations during the whole 
observing period were less than 10 $\%$.  

\placetable{tbl-1}

\section{Results} 
\subsection{CH$_{\rm 3}$OH and C$^{\rm 34}$S 
Spectra toward IRAM 04191}

Figure 1 shows profiles of the CH$_{\rm 3}$OH and C$^{\rm 34}$S lines 
toward the central position of IRAM 04191. Since the integration time 
toward the central position was much longer (37 min) than that in other 
mapping positions, signal-to-noise ratios of these spectra (rms $\sim$ 
0.025 K) are better than those at other points. Three transitions of
CH$_{\rm 3}$OH, 
i.e., 2$_{\rm 0}$--1$_{\rm 0}$ A$^{\rm +}$, 2$_{\rm -1}$--1$_{\rm - 
1}$ E, and 2$_{\rm 0}$--1$_{\rm 0}$ E were detected at the position
of the protostar, while the 2$_{\rm 1}$--1$_{\rm 1}$ E 
transition was not detectable at any observed positions,
including the stellar position, at 1 $\sigma$ noise 
level of $\sim$ 0.06 K. The C$^{\rm 34}$S line was clearly detected 
toward the protostellar position, with intensity less than half of that 
of the CH$_{\rm 3}$OH (2$_{\rm 0}$--1$_{\rm 0}$ A$^{\rm +}$) line.

Both the CH$_{\rm 3}$OH and C$^{\rm 34}$S lines at the position of IRAM 
04191 show single-peaked, Gaussian-like profiles with possible redshifted 
wings. Gaussian fittings to these profiles show that peak velocities 
and line widths of the spectra are $\sim$ 6.6 - 6.7 km s$^{\rm -1}$ and 
$\sim$ 0.4 - 0.5 km s$^{\rm -1}$, respectively (see Table 2). 
The C$^{\rm 34}$S profile is essentially same as in Belloche et al. 
(2002). Andr\'e et 
al. (1999a) have shown that
CS ($J$=2--1) and H$_{\rm 2}$CO (2$_{\rm 
12}$-- 1$_{\rm 11}$) lines from IRAM 04191 exhibit the  
double-peaked, asymmetric spectral signature of protostellar collapse 
\cite{zho92,mye95}. The peak velocity of the CH$_{\rm 3}$OH and 
C$^{\rm 34}$S lines corresponds to the velocity of the absorption 
``dip'' of the CS 
($J$=2--1) and H$_{\rm 2}$CO (2$_{\rm 12}$--1$_{\rm 11}$) profiles, 
suggesting that the CH$_{\rm 3}$OH and C$^{\rm 34}$S lines are 
optically thin. Hereafter, we refer to 6.7 km s$^{\rm -1}$ as a systemic 
velocity of IRAM 04191, which is shown in a vertical dotted line in Figure 1.

\placefigure{fig1} 

\subsection{Spatial and Velocity Distribution of 
the CH$_{\rm 3}$OH Emission}

In this subsection, spatial and velocity distribution of 
the CH$_{\rm 3}$OH emission is discussed. We only adopt the 
CH$_{\rm 3}$OH (2$_{\rm 0}$--1$_{\rm 0}$ A$^{\rm +}$) line here 
because the 2$_{\rm -1}$--1$_{\rm - 1}$ E transition shows essentially 
the same distribution as that the 2$_{\rm 0}$--1$_{\rm 0}$ A$^{\rm +}$ 
transition shows in our mapping observations, and because 
the 2$_{\rm 0}$--1$_{\rm 0}$ E transition, which has a higher 
energy level than the other two detected lines (see Table 1), 
is detected only at several positions in the mapping observations. 
Figure 2 shows velocity channel maps of 
the CH$_{\rm 3}$OH (2$_{\rm 0}$--1$_{\rm 0}$ A$^{\rm +}$) line with 
a velocity interval of 0.2 km s$^{\rm -1}$. 
The emission was significantly 
detected within the LSR velocity range from 5.0 to 7.4 km s$^{\rm -1}$. At 
blueshifted velocities ranging from 5.1 to 6.1 km s$^{\rm -1}$, there 
are two CH$_{\rm 3}$OH emission features detected: one appears 
$\sim$ 60$\arcsec$ south to the central star and the other at $\sim$ 
90$\arcsec$ southwest to the star. The peak positions of these two 
emission features do not change much in this velocity range. At 
velocities from 6.3 to 7.3 km s$^{\rm -1}$, around the systemic velocity, 
another CH$_{\rm 3}$OH emission feature was detected in the vicinity 
of IRAM 04191. The peak position of this emission feature shifts from 
the northwest (at 6.3, 6.5 and 6.7 km s$^{\rm -1}$) to the southeast (at 
6.9 and 7.1 km s$^{\rm -1}$), and then to the northeast of the central 
star (at 7.3 km s$^{\rm -1}$). At redshifted velocities 
(6.9 - 7.3 km s$^{\rm -1}$), there is an 
additional CH$_{\rm 3}$OH emission feature detected at $\sim$ 
60$\arcsec$ west to the central star. This feature is likely 
to be independent of the 6.3 - 6.7 km s$^{\rm -1}$ component seen at the 
northwest, since the line profile toward this feature shows a double peak, 
one from the 6.3 - 6.7 km s$^{\rm -1}$ component and the other from this 
new feature (see Figure 3).
At the northeastern edge of the 
mapping region, there appears strong CH$_{\rm 3}$OH emission but the 
limited observing time prevented us from extending our mapping region 
to cover this component. Hence, this emission will not be discussed in 
this paper. We note that toward IRAS 04191+1523 (filled squares in 
Figure 2) the CH$_{\rm 3}$OH emission distribution seems to be local 
minimum and there is no clear CH$_{\rm 3}$OH condensation associated 
with IRAS 04191+1523.

The above examination of the CH$_{\rm 3}$OH channel maps shows that four 
CH$_{\rm 3}$OH emission features at different velocities are 
present in IRAM 04191. In order to investigate the nature of these 
different emission features in more detail, the channel maps are 
integrated over three different velocity ranges, i.e., 5.0 - 6.5 km 
s$^{\rm - 1}$, 6.5 - 6.9 km s$^{\rm -1}$, and 6.9 - 7.4 km s$^{\rm -1}$, 
which roughly distinguish the different CH$_{\rm 3}$OH emission 
components from each other. Figure 3 shows three maps integrated over 
the above velocity ranges, superposed on the CO ($J$=2--1) outflow 
map obtained by Andr\'e et al. (1999a).

The map of the systemic velocity (Figure 3; 6.5 - 6.9 km s$^{\rm -1}$) 
shows a CH$_{\rm 3}$OH condensation elongated in the 
east-west direction (P.A. $\sim$ 110$^{\circ}$), which is almost 
perpendicular to the outflow axis.  
The southern side of the condensation shows a concavity, 
through which the blueshifted outflow extends to south. This 
condensation is most likely a molecular envelope associated with 
IRAM~04191, although the 
position of IRAM~04191 is slightly ($\sim$ 20$\arcsec$ east) offset 
from the center of the condensation. This kind of the elongated envelope
is often associated with protostars 
\cite{miz94,zho94,oh97a,oh97b,sai99}.
Hence, hereafter we call this 
condensation ``ENV''.
The deconvolved size and the deconvolved line 
width of ENV were measured to be $\sim$ 0.07 pc $\times$ 
0.04 pc and $\sim$ 0.5 km s$^{\rm -1}$, respectively. These values are 
comparable to the typical size and line width of envelopes associated 
with low-mass protostellar sources 
\cite{ben89,miz94,zho94}. 

As has been suggested by the channel maps in Figure 2, 
there is a velocity gradient 
along the major axis of ENV. Figure 4, the position-velocity 
(P-V) diagram cutting along the major axis, demonstrates the velocity 
gradient clearly, as indicated by the thick dashed line.
This velocity gradient is most probably attributed to rotation of ENV
because it is obviously seen along the major axis of ENV.
This velocity gradient was also seen in the 
N$_{\rm 2}$H$^{+}$ ($J$=1--0) line (Belloche et al. 2002).
It should be 
noted that, the velocity gradient can be seen only in the vicinity 
of the central star 
(within $\sim$ 0.01~pc from the central star), while 
the outer part of ENV, particularly the western side, does 
not show a clear velocity shift from the systemic velocity.
This might be because the rotation of ENV is differential,
so that the rotation velocity becomes larger in the inner part of ENV.
The velocity gradient was measured 
to be $\sim$ 40~km~s$^{\rm -1}$~pc$^{\rm -1}$, which is 10-100 
times larger than that in other molecular cores measured by Goodman 
et al. (1993). This difference may result from the difference in the 
radius where velocity gradients are measured, i.e., in our case the 
gradient was measured at $\sim$ 0.01~pc in radius, while Goodman et 
al. (1993) measured those gradients typically at $\sim$ 0.1~pc in 
radius.

The map integrated over 5.0 - 6.5 km s$^{\rm -1}$ (the middle panel in 
Figure~3) shows two compact ($\sim$ 0.03 pc) condensations,
connected by extended emission at lower levels. 
These condensations correspond to the two emission features 
seen at blueshifted 
velocities in the channel maps.
Hereafter, we will call 
these compact CH$_{\rm 3}$OH condensations ``BLUE1'' (the western, 
stronger one) and ``BLUE2'', respectively. 
These two blueshifted gas components were also 
detected 
in the CS(2--1) line (Belloche et al. 2002; see their Fig. 6). 
As described below, physical 
properties of BLUE1 and BLUE2 are significantly 
different from those of ENV: 
\begin{itemize} \item[(1)] BLUE1 and BLUE2 are not 
located in the vicinity of the central star, but located at the southern 
edge of ENV. Comparison with the CO outflow 
map shows that BLUE1 and BLUE2 locate on the the western and 
eastern sides of the blueshifted outflow lobe, respectively. 
Specifically, the peak of BLUE1 locates just beside the strongest 
peak in the blueshifted CO outflow. 
\item[(2)] Line profiles toward 
BLUE1 and BLUE2 show non-Gaussian shapes, with much wider line widths 
(up to 2.0 km s$^{\rm -1}$) than that of ENV and 
centroid velocities blueshifted by $\sim$ 0.8 km s$^{\rm -1}$ from the 
systemic velocity (Table 2, 3). The P-V diagram (Figure 5) of the 
CH$_{\rm 3}$OH emission along the blueshifted outflow 
(solid line in Figure 3) shows abrupt increase of the line width 
from $\sim$ 0.5 km s$^{\rm -1}$ to $\sim$ 2.0 km s$^{\rm -1}$ at the 
position of BLUE 1. \end{itemize} These results suggest that 
BLUE1 and BLUE2 are physically different from ENV. The 
origin of BLUE1 and BLUE2 will be discussed in $\S$4 in detail.

In the map integrated over 6.9 - 7.4 km s$^{\rm -1}$ (the lower panel in 
Figure~3), there is a CH$_{\rm 3}$OH condensation detected to the west 
of the protostar, as well as a condensation appearing to the east of the 
protostar. 
As seen in Figure 2, the eastern condensation seems to be a 
part of ENV, while it delineates the 
southeast side of the redshifted outflow and could represent the 
interaction region between ENV and the redshifted outflow \cite{bel02}. 
On the other hand, the 
western condensation is not a part of ENV, but 
the one corresponding to the emission feature detected at redshifted 
velocities in the channel maps. Hereafter, we will call the redshifted 
western condensation ``RED''. The line profile toward RED shows a 
double-peaked feature, i.e., one peaked at 6.7 km s$^{\rm -1}$ arising 
from ENV, and the other peaked at 7.1 km s$^{\rm -1}$ arising 
from RED. It is 
found that, while the size of RED is as compact as that of BLUE1 
or BLUE2 ($\sim$ 0.03 pc), the line width is narrow 
($\sim$ 0.3 km s$^{\rm -1}$).

\placefigure{fig2} 
\placefigure{fig3} 
\placefigure{fig4} 
\placefigure{fig5} 
\placetable{tbl-2}

\subsection{Spatial and Velocity Distribution of the C$^{\rm 34}$S 
Emission}

Figure 6 shows velocity channel maps of the C$^{\rm 34}$S 
($J$=2--1) emission. 
The velocity
range and interval of Figure 6 are same as those of the 
CH$_{\rm 3}$OH (2$_{\rm 0}$--1$_{\rm 0}$ A$^{+}$) channel maps 
shown in Figure 2.
The typical intensity of the C$^{\rm 34}$S emission is a
factor of 2 lower than the intensity of the CH$_{\rm 3}$OH emission in 
the mapped area. The 
C$^{\rm 34}$S emission was not detected at every velocity channel 
where the CH$_{\rm 3}$OH emission was detected, but only in the 
velocity range 
from 6.3 to 7.1 km s$^{\rm -1}$ where ENV and a part of RED 
were detected in the CH$_{\rm 3}$OH map.
In that velocity range, 
the spatial distribution of the C$^{\rm 34}$S emission is 
similar to that of the CH$_{\rm 3}$OH emission. Since the 
C$^{\rm 34}$S emission is weaker than the CH$_{\rm 3}$OH 
emission, we will concentrate on the CH$_{\rm 3}$OH data in the following 
discussions.

\placefigure{fig6} 

\subsection{CH$_{\rm 3}$OH Abundance}

The molecular abundance is one of the most important parameters to 
describe chemical properties of the four condensations, 
that is, ENV, BLUE1, BLUE2, and RED. 
We have estimated the 
molecular abundance of CH$_{\rm 3}$OH for each condensation 
by dividing the CH$_{\rm 3}$OH column density ($N_{\rm 
CH_3OH}$) by the molecular hydrogen column density ($N_{\rm H_2}$), 
both of which are measured at 
the central position of each condensation. 
In order to estimate $N_{\rm CH_3OH}$, we assume that the 
emission lines are optically thin, which is valid for the emission lines 
we observed (see Table 3). In addition we assume Local Thermal Equilibrium 
(LTE) condition of the CH$_{\rm 3}$OH transitions 
and the equal partition function for both A-state and E-state CH$_{\rm 3}$OH 
\cite{fri88}. 
Under these conditions, the column density is given by the 
following formula; \begin{equation} N_{\rm 
CH_3OH}=\frac{8\pi\nu^{3}}{c^{3}}\frac{1}{(2J_l+3)A} \frac{Z({\rm 
T_{ex}})}{\exp(-\frac{E_l}{k{\rm T_{ex}}}) (1-\exp(-\frac{h\nu}{k{\rm 
T_{ex}}}))} \frac{{\rm T_B^*}\Delta{\rm V}}{J({\rm T_{ex}})-J({\rm 
T_{bg}})}, \end{equation} where \begin{equation} J({\rm 
T})=\frac{\frac{h\nu}{k}} {\exp(\frac{h\nu}{k{\rm T}})-1}. 
\end{equation} In the above expressions, $h$ is the Planck constant, $k$ 
is the Boltzmann constant, $c$ is the speed of light, $\nu$ is the line 
frequency, T$_{\rm ex}$ is the excitation temperature of the CH$_{\rm 
3}$OH transitions, ${\rm T_{bg}}$ is the background radiation 
temperature, $A$ is the Einstein A coefficient of the transition, $Z$ is 
the partition function, $E_l$ is the rotational energy level of the lower 
energy state, $J_l$ is the rotational quantum number of the lower energy 
state, and ${\rm T_B^*}\Delta$V is the integrated line 
intensity. We adopted T$_{\rm ex}$ of 10 K for all the condensations, 
according to our estimation of T$_{\rm ex}$ from 
excitation analyses using the data of the 
($J_{\rm K}$=5$_{\rm K}$--4$_{\rm K}$) band obtained 
with the JCMT and the data of the 
($J_{\rm K}$=2$_{\rm K}$--1$_{\rm K}$) band presented here
on the assumption of the Large Velocity Gradient (LVG) condition 
\cite{gol74,sco74}. 
Details of the LVG analyses with the 
multi-transitional CH$_{\rm 3}$OH lines will be presented in a 
subsequent paper. 

The column densities of molecular hydrogen, on the other hand, were 
estimated using the 1.3 mm dust continuum data of IRAM~04191 taken 
with the IRAM 30~m telescope \cite{an99b,mot01}. We read out 
intensity values of the dust continuum emission at each central position, and 
calculated $N_{\rm H_{2}}$ by using the formula and the dust 
temperature used in Andr\'e et al. (1999a) and Motte \& Andr\'e (2001).
CH$_{\rm 3}$OH abundances were estimated from 
these $N_{\rm CH_3OH}$ and $N_{\rm H_2}$. Here, we did not take into account 
the difference of the beam size between our observations 
(17$\arcsec$) and the 1.3 mm continuum observations with the IRAM 30 
m telescope (13$\arcsec$). All the derived values are listed in 
Table~3.

The CH$_{\rm 3}$OH abundance toward the protostar was estimated to 
be $\sim$ 2.1 $\times$ 10$^{\rm -9}$, which is indistinguishable from 
the CH$_{\rm 3}$OH abundance toward the cyanopolyyne peak in TMC-1 
($\sim$ 2.0 $\times$ 10$^{\rm -9}$ ; Pratap et al. 1997, Takakuwa et 
al. 2000). 
On the other hand, the CH$_{\rm 3}$OH 
abundances toward BLUE1, BLUE2, and RED are 5-10 times 
larger than that toward the protostar. We will discuss this abundance 
enhancement later in more detail.

\subsection{Physical Properties of the CH$_{\rm 3}$OH Condensations}

In addition to the chemical properties of the CH$_{\rm 3}$OH 
condensations discussed above, we examined the physical properties of 
the condensations. 
The LTE mass of each condensation, M$_{\rm LTE}$,
was calculated using the column density and
size listed in Table~3. 
The fractional abundance of CH$_{\rm 3}$OH 
for each condensation we estimated in the previous
section (also listed in Table~3) was used to estimate the total 
H$_2$ gas mass. 
The LTE mass obtained in this way is 2.3 M$_{\odot}$ for
ENV, while less than 0.1 M$_{\odot}$ for the other three condensations.
In order to examine whether or not the CH$_{\rm 3}$OH condensations are 
gravitationally bound, we
compared M$_{\rm LTE}$ with the virial mass (M$_{vir}$) 
calculated from the size (D) 
and the FWHM line width ($\Delta$V)
using the following formula,
\begin{equation} {\rm M}_{vir} = 
\frac{5DC_{eff}^{\rm 2}}{2G}, \end{equation} where \begin{equation} 
C_{eff}^{2} \sim \frac{\Delta{\rm V^{2}}}{\rm 8ln2}. \end{equation} In 
the above expressions $C_{eff}$ is the effective sound speed of gas and 
$G$ is the gravitational constant. Note that the above expressions 
assume a condensation with a uniform density distribution. 
The masses of each CH$_{\rm 3}$OH condensation derived from 
these two different ways are listed in Table 3.

The virial mass of ENV is similar to 
its LTE mass, suggesting that ENV is probably virialized. In 
addition, the derived mass is comparable to the typical mass of the 
low-mass star-forming envelopes \cite{miz94,zho94}. These results 
suggest that ENV is an envelope being in the course of star-formation,
and ENV is indeed associated with the protostar of IRAM~04191.

In contrast to ENV, the virial masses of BLUE1 and BLUE2 are 
$\sim$ 30-100 times larger than their LTE masses,
suggesting that the 
blueshifted components are most likely gravitationally unbound. 
Similarly to the blueshifted components, the virial mass of RED is 
$\sim$10 times larger than its LTE mass, 
which suggests that RED is also likely gravitationally unbound.

\placetable{tbl-3}

\section{Discussion}
\subsection{The Origin of the Blueshifted and Redshifted CH$_{\rm 3}$OH 
Condensations}

Our CH$_{\rm 3}$OH observations of IRAM 04191 have revealed that, in 
addition to the envelope (ENV) associated with star formation activity, 
there are two blueshifted gas condensations (BLUE1, BLUE2) and one 
redshifted gas condensation (RED). 
The observed properties of the blueshifted components, i.e., 
located at the periphery of the blueshifted CO outflow, the wider 
line width, and the centroid velocity blueshifted from the systemic 
velocity, suggest that the blueshifted components are likely 
to be related to the blue lobe of the CO outflow. 
The total momentum of the blueshifted components is comparable to the CO 
outflow momentum ($\sim$ 0.12~M$_{\odot}$ km s$^{-1}$; Andr\'e et 
al. 1999a), suggesting that the blueshifted components can be pushed by 
the outflow. In addition, offset velocities 
(V$_{off}$ $\equiv$ (V$_{\rm CH_{3}OH}$ -- V$_{sys}$) / cos $i$; 
where $i$ is the inclination angle of the outflow
from the line of sight) of the 
blueshifted CH$_{\rm 3}$OH condensations are comparable to 
velocities necessary for the blueshifted condensations 
to escape from the gravitational potential of 
the central star-forming system, that is, 
V$_{g}$ $\equiv$ $\sqrt{\frac{2G{\rm M}}{\rm R}}$, 
where M is the mass of the envelope ($i.e.,$ = 2.3 
M$_{\odot}$) + star (0.05 M$_{\odot}$) and R is the projected distance 
from the protostar (Table 4). 
These results suggest that both BlUE1 and BLUE2 
are most likely to be formed in consequence of interaction between the 
outflow and the ambient dense gas surrounding the protostar, and that 
the blueshifted components are pushed away from the ambient cloud core. 
The eastern condensation, which seems to be a part of ENV 
and delineates the southeast side of the redshifted outflow 
(Figure 3 lower panel), might also represent a core 
interacting with the redshifted outflow \cite{bel02}. 

Nakano, Hasegawa, \& Norman (1995) theoretically investigated 
mechanisms of setting the final stellar mass in the process of star 
formation, and proposed that the dispersal of the parent cloud core as a 
result of the interaction with the mass outflow (e.g. Momose et al. 
1996; Velusamy \& Langer 1998) is crucial for fixing the final 
stellar mass. 
Hence, it is interesting to discuss whether the 
interaction with the IRAM~04191 molecular outflow plays an important 
role in the dissipation of the surrounding molecular cloud core and the 
process of the stellar mass determination. As discussed in 
$\S$3.5, the blueshifted components are gravitationally unbound, 
suggesting that they will most likely dissipate eventually. The time 
scale of the dissipation, roughly estimated as $D/C_{eff}$ where $D$ is 
the size of the condensation and $C_{eff}$ is the effective sound speed 
derived from the line width, is $\sim$ 5 $\times$ 10$^{\rm 4}$ yr. This 
time scale is comparable to the typical life time of the accreting 
protostar, $\sim 10^5$~yr \cite{ada87,ada90,lad91}. 
Hence, the 
dissipation of the surrounding cloud core may sufficiently proceed 
around IRAM~04191 during the main accretion phase, resulting in 
setting the stellar mass.

We note, however, that the total mass of the blueshifted components 
($\sim$ 0.15 M$_{\odot}$), which will most likely dissipate, is at most 
7$\%$ of the total mass of the envelope, suggesting that the total 
amount of the dissipating gas may not be high enough to have a significant 
impact onto the supply of matter for the mass accretion. Since 
IRAM~04191 is considered to be a very young class 0 protostar 
\cite{an99a,an99b}, the outflow may just have started interacting with 
the surrounding cloud core so that the core dissipation may not proceed 
enough yet. We speculate that the outflow will keep interacting with 
the surrounding cloud core, resulting in dissipation of significant 
fraction of the surrounding cloud core.

The origin of the 
redshifted component, on the other hand, is less clear. Its narrow 
line width ($\sim$ 0.3~km s$^{\rm -1}$) suggests that the redshifted 
component is not related to the outflow. There is no clear star 
formation activity found in the redshifted condensation. The physical
properties of the redshifted condensation, i.e., the narrow line width, 
the small mass ($\sim$ 0.07 M$_{\odot}$) 
and size ($\sim$ 0.02 pc), and non-virialization, are rather similar to 
those of 
small-scale clumps found in starless cores \cite{lan95,pen98,oha99}. 

\placetable{tbl-4}

\subsection{CH$_{\rm 3}$OH Abundance Enhancement}

Previous observations of molecular outflows associated with L1157 
\cite{bac95,ave96,bac01}, NGC1333 IRAS 2 \cite{san94,bac98}, 
and BHR 71 \cite{gar98} have also showed 
CH$_{\rm 3}$OH emission arising from regions where the outflows 
interact with 
the surrounding dense gas. An important, common result obtained 
from these previous observations is that the CH$_{\rm 3}$OH emission 
arising from the interaction regions is much more intense and broader 
than that from the positions of the driving sources. 
The more intense emission has been attributed to
bowshock heating, caused by the interaction between the outflows and
the dense gas, 
leading to the enhancement of the CH$_{\rm 3}$OH abundance (see Table 5). 
Is the blueshifted CH$_{\rm 3}$OH emission in IRAM~04191 under
influence of such bowshock heating?

As discussed in $\S$3.4., the CH$_{\rm 3}$OH abundance in the blueshifted 
condensations is enhanced by a factor of $\sim$ 10 
with respect to that at the position of the protostar. This abundance 
enhancement might be due to shocks associated with the interaction 
between the blueshifted outflow and the surrounding dense gas. However, the 
enhancement seen in the blueshifted condensations 
is not as dramatic as that seen in 
the above-mentioned three outflows associated with shocks: the CH$_{\rm 
3}$OH abundances in these outflows with shocks were estimated to be 
enhanced by more than 2 orders of magnitude 
\cite{san94,bac95,bac98,bac01,ave96,gar98}. We also note that the 
C$^{\rm 34}$S emission, which is also enhanced in the shocked regions 
\cite{san94,bla95,gar98,bac01} was barely detected at the positions of 
the blueshifted components in IRAM 04191. 

If CH$_{\rm 3}$OH is enhanced due to shocks caused by the 
outflow, then the CH$_{\rm 3}$OH enhancement should be 
localized on the shocked regions.
In order to see if there is such localization of the
CH$_{\rm 3}$OH enhancement, the CH$_{\rm 3}$OH abundance map in 
the IRAM~04191 region, shown in Figure 7, was made 
from the ratio of 
the total integrated intensity map of the CH$_{\rm 3}$OH emission to 
the 1.3 mm dust continuum map \cite{an99a,mot01}. We adopted the same 
assumptions as those in $\S$3.4 to create this abundance map. We can 
clearly see that the CH$_{\rm 3}$OH abundance is extensively enhanced 
in the periphery of the entire region, including the position of RED, 
which shows no sign of interaction. 
There is no local enhancement of 
the CH$_{\rm 3}$OH abundance toward BLUE2 and probably BLUE1, 
for which the fractional abundance was not determined because of 
less signal-to-noise ratio of the 1.3 mm dust continuum map \cite{mot01}. 
We have 
also conducted sub-millimeter CH$_{\rm 3}$OH ($J_{\rm K}$=5$_{\rm 
K}$--4$_{\rm K}$, 7$_{\rm K}$--6$_{\rm K}$) observations with the 
JCMT toward BLUE1, and have found the temperature of the 
blueshifted component is cold ($\sim$ 10 K), suggesting that there is 
no local shock heating toward the blueshifted component (our 
forthcoming paper).
These results suggest that the 
CH$_{\rm 3}$OH abundance enhancement seen in the blueshifted 
components is not due to shocks associated with the interaction 
between the blueshifted outflow and the dense core. 

We consider that the observed abundance gradient is likely to reflect 
the chemistry in the protostellar core 
rather than the effect of shocks due to the 
outflow. Recently, Aikawa et al. (2001) have constructed chemical 
evolutionary models of a collapsing dense core. They found that 
chemical abundances of molecules are functions of both the 
evolutionary time-scale and the distance from the core center. 
They showed that, 
the abundances of ``early-type'' molecules, which are 
considered to be more abundant in the early stage of chemical evolution of 
dense cores, significantly decrease in the central regions of the dense core
as the evolutionary time goes on, resulting in abundances
of early-type molecules appearing ``enhanced''
in the periphery of the dense core. Since 
CH$_{\rm 3}$OH is considered to be one of the early-type 
molecules \cite{pra97,tak98,tak00}, 
the observed CH$_{\rm 3}$OH 
abundance variation in IRAM 04191 is consistent with their model.
Such abundance gradients of early-type
molecules such as CCS were
also shown in other observations \cite{kui96,oha99,oha00}. 

It is interesting to discuss why the blueshifted CH$_{3}$OH emission 
in IRAM~04191 is not due to shock heating by the blueshifted 
outflow, which most probably interacts with the ambient cloud. In 
Table 5, dynamical properties of the CO outflow from IRAM~04191 are 
compared with those of the three outflows with shock-enhanced 
CH$_{\rm 3}$OH emission. 
The table shows that the flow velocities (V$_{\rm flow}$) of 
the CO outflows are 
very different 
between the three CO outflow and IRAM~04191 outflow. The flow 
velocities of the three CO outflows in the shocked regions are all above 
20 km s$^{\rm -1}$, whereas the flow velocity of IRAM~04191 outflow 
is below 10 km s$^{\rm -1}$. 
We also note that the 
line widths and the offset velocity (V$_{off}$) 
of the CH$_{\rm 3}$OH components in IRAM 
04191 are also small as compared to those in the other shocked outflows 
(Table 5). 
Shocks associated with molecular outflows 
from protostars are generally considered to be C-type shocks 
\cite{hol97}. 
According to detailed models of C-type shocks including the calculation 
of the chemical composition (i.e., coolants) and the grain size 
distribution, heating caused by C-type shocks is a strong function of 
the flow velocity, or the shock velocity \cite{ka96a,ka96b,ber98}. 
These models show that shock velocity higher than $\sim$ 10 km 
s$^{\rm -1}$ is necessary to yield temperature enhancement higher than 
120 K, which is considered to be the minimum required temperature to 
release CH$_{\rm 3}$OH on dust grains into gas phase \cite{san93}.
Thus, the difference in V$_{\rm flow}$ may explain why 
IRAM~04191 is not associated with CH$_{\rm 3}$OH emission enhanced 
due to shocks. 

\placefigure{fig7} 
\placetable{tbl-5}

\section{Conclusions}

We have carried out mapping observations around the young Class 0 
protostar, IRAM 04191+1522, in the CH$_{\rm 3}$OH ($J_{\rm 
K}$=2$_{\rm K}$--1$_{\rm K}$) and C$^{\rm 34}$S ($J$=2--1) lines 
with the NRO 45 m telescope. The main results are as follows.

1. A dense gas envelope surrounding the protostar was traced 
both in the CH$_{\rm 3}$OH and C$^{\rm 34}$S lines. 
The envelope has a size of $\sim$ 0.07 pc $\times$ 0.04 pc, line 
width of $\sim$ 0.5 km s$^{\rm -1}$, and LTE mass of 
$\sim$ 2.3 M$_{\odot}$. 
There is a steep velocity gradient of $\sim$ 
40 km s$^{\rm -1}$ pc$^{\rm -1}$ along the NW-SE direction, i.e., 
perpendicular to the axis of the outflow, in the envelope. 
This velocity gradient is interpreted
as the rotation of the envelope.

2. The CH$_{\rm 3}$OH map shows two compact ($\sim$ 0.03 pc in size), 
blueshifted ($\sim$ 0.8 km s$^{\rm -1}$ from the systemic velocity) 
condensations with wide line widths (up to 2.0 km s$^{\rm -1}$) 
at the periphery of the blueshifted lobe of the CO outflow. 
The compact condensations have momenta 
($\sim$ 0.06 M$_{\odot}$ km s$^{\rm -1}$) comparable to that of
the blueshifted molecular outflow. 
These results suggest that 
the two blueshifted components are likely to be 
formed by means of interaction between the outflow and its
surrounding dense gas. 
The LTE masses of these blueshifted components ($\sim$ 0.07
M$_{\odot}$) are two orders of magnitude lower than their virial masses, 
suggesting that these condensations are not bound by their self-gravity, 
and most likely will dissipate eventually.

3. The CH$_{\rm 3}$OH abundance in the two blueshifted condensations is 
enhanced by a factor of $\sim$ 10 with respect to that toward the 
position of 
the protostar ($\sim$ 2.1 $\times$ 10$^{\rm -9}$). 
However, this 
abundance enhancement is less significant as compared to that seen in the
outflows with shock heating. In addition, the spatial variation of the
CH$_{\rm 3}$OH abundance does not show the local enhancement at the 
positions of the blueshifted condensations, but shows the trend of 
decrease toward the center and increase toward the periphery. 
Such an abundance variation is explained 
by the effect of chemical evolution of the collapsing dense core rather
than by the shock chemistry.

4. A comparison of the properties of the CO outflow in IRAM 04191 with
those with shock heating suggests that the velocity of the outflow in IRAM
04191 (less than 10 km s$^{\rm -1}$) is not high enough to produce shocks 
that can release the CH$_{\rm 3}$OH molecules from dust mantles.

\acknowledgments 
We are grateful to M. Choi for his fruitful 
discussions. We would like to thank all the NRO staff supporting this 
work. S.T. was supported by a postdoctoral fellowship of the Institute of 
Astronomy and Astrophysics, Academia Sinica. N.O. is supported in part 
by NSC grant 91-2112-M-001-029.

\clearpage

\figcaption[fig1.ps]{CH$_{\rm 3}$OH 
($J_{\rm K}$=2$_{\rm 0}$--1$_{\rm 0}$ A$^{\rm +}$, 
2$_{\rm -1}$--1$_{\rm -1}$ E, 2$_{\rm 0}$--1$_{\rm 0}$ E, and 
2$_{\rm 1}$--1$_{\rm 1}$ E) and C$^{\rm 34}$S ($J$=2--1) line profiles 
toward the protostellar position of IRAM 04191+1522. The vertical 
dotted line indicates the systemic velocity of 6.7 km s$^{\rm -1}$.
\label{fig1}}

\figcaption[fig2.eps]{Velocity channel maps of the CH$_{\rm 3}$OH 
($J_{\rm K}$=2$_{\rm 0}$--1$_{\rm 0}$ A$^{\rm +}$) emission in IRAM 
04191+1522. The channel widths are 0.2 km s$^{\rm -1}$ and the 
center velocities are shown in the Figure. 
Contour levels are from 0.12 
K by steps of 0.08 K. Cross marks indicate the position of the 
protostar, IRAM 04191+1522. Filled squares shows the position of 
IRAS 04191+1523. Dot marks are the observed points.\label{fig2}}

\figcaption[fig3.eps]{Velocity channel maps of the CH$_{\rm 3}$OH 
($J_{\rm K}$=2$_{\rm 0}$--1$_{\rm 0}$ A$^{\rm +}$) emission in three 
velocity regimes (Systemic : 6.5 - 6.9 km s$^{\rm -1}$, BLUE : 5.0 - 6.5  
km s$^{\rm -1}$, and RED : 6.9 - 7.4 km s$^{\rm -1}$) plus line profiles 
toward the representative positions (ENV, BLUE1, BLUE2, and RED) 
in IRAM 04191+1522. Contour levels in the channel maps are from 0.10 
K km s$^{\rm -1}$ by steps of 0.04 K km s$^{\rm -1}$. In the channel 
maps we overlay the map of the CO outflow by Andr\'e et al. (1999a) 
(Red contours indicate red lobe and blue contours blue lobe of the CO 
outflow). Marks in the channel maps have the same meanings as in 
Figure 2. Straight lines in the channel maps indicate the cut of the 
position-velocity diagram shown in Figure 5. 
In the line profiles we show the systemic velocity of 
6.7 km s$^{\rm -1}$ by a dashed vertical line. \label{fig3}}

\figcaption[fig4.ps]{A position-velocity (P-V) diagram of the 
CH$_{\rm 3}$OH ($J_{\rm K}$=2$_{\rm 0}$--1$_{\rm 0}$ A$^{\rm +}$) emission 
along the major axis of the protostellar envelope (P.A.=120$^{\circ}$) in 
IRAM 04191+1522. Contour levels are from 0.12 K by steps of 0.08 K. A 
vertical dotted line in the figure indicates the systemic velocity 
(= 6.7 km s$^{\rm -1}$), and a horizontal line in the figure represents 
the position of the protostar. A bold dotted line indicates the velocity 
gradient of the protostellar envelope around the protostar.\label{fig4}}

\figcaption[fig5.ps]{A P-V diagram of the CH$_{\rm 3}$OH 
($J_{\rm K}$=2$_{\rm 0}$--1$_{\rm 0}$ A$^{\rm +}$) emission along the cut 
which crosses BLUE1 with P.A.=30$^{\circ}$ (straight lines in Figure 
3). Contour levels are from 0.15 K by steps of 0.10 K. We can see 
sudden increase of the line width at the position of BLUE 1.
\label{fig5}}

\figcaption[fig6.eps]{Same as Figure 2 but for the C$^{\rm 34}$S 
($J$=2--1) emission.\label{fig6}}

\figcaption[fig7.eps]{A map of the CH$_{\rm 3}$OH abundance derived 
from the ratio of the total integrated intensity map of the 
CH$_{\rm 3}$OH ($J_{\rm K}$=2$_{\rm 0}$--1$_{\rm 0}$ A$^{\rm +}$) 
line to the 1.3 mm dust continuum map by Motte \& Andr\'e (2001) in 
IRAM 04191+1522. The innermost contour corresponds to the 
CH$_{\rm 3}$OH abundance of 1.6 $\times$ 10$^{\rm -9}$, and the 
abundance increases by a factor of 2, 3, 4, 10 toward the outer 
contours as indicated in the figure. In the figure we plot the position 
of IRAS 04191+1523 (north-eastern), IRAM 04191+1522 (central), 
BLUE1 (south-western ; outside the border of the map), BLUE2 
(south of IRAM 04191+1522), and RED (west of IRAM 04191+1522) 
by cross marks. The signal-to-noise ratio of the 1.3 mm dust 
continuum map toward BLUE1 is worse than the other positions and 
therefore BLUE1 is outside the map boundary. We can see that the 
abundance systematically increases in the outer part of the 
protostellar envelope. \label{fig7}}

\clearpage 
\begin{deluxetable}{lllcl} 
\tablecaption{Observed Molecular Lines\label{tbl-1}} 
\tablewidth{0pt} 
\tablehead{\colhead{Molecule} & 
\colhead{Transition} & \colhead{Frequency} & 
\colhead{Einstein A Coefficient} & \colhead{E$_{u}$} \\ \colhead{} & \colhead{} & 
\colhead{(GHz)} & \colhead{(s$^{\rm -1}$)} & \colhead{(K)} }
\startdata
CH$_{\rm 3}$OH\tablenotemark{a} &$J_{\rm K}$ = 2$_{\rm 0}$--1$_{\rm 0}$ A$^{\rm +}$ 
&96.74142  &3.41 $\times$ 10$^{\rm -6}$ &6.96\tablenotemark{c}\\
&$J_{\rm K}$ = 2$_{\rm -1}$--1$_{\rm -1}$ E  &96.73939  &2.56 $\times$ 10$^{\rm -6}$ 
&4.64\tablenotemark{c}\\
&$J_{\rm K}$ = 2$_{\rm 0}$--1$_{\rm 0}$ E   &96.74458  &3.41 $\times$ 10$^{\rm -6}$ 
&12.2\tablenotemark{c}\\
&$J_{\rm K}$ = 2$_{\rm 1}$--1$_{\rm 1}$ E   &96.75551  &2.62 $\times$ 10$^{\rm -6}$ 
&20.1\tablenotemark{c}\\ 
\hline
C$^{\rm 34}$S\tablenotemark{b}  &$J$ = 2--1 &96.412982 &1.60 $\times$ 10$^{\rm -5}$ 
&6.93\\ 
\enddata
\tablenotetext{a}{Line parameters of CH$_{\rm 3}$OH are from 
Xu \& Lovas (1997).}
\tablenotetext{b}{Line 
parameters of C$^{\rm 34}$S are from Swade (1989).} 
\tablenotetext{c}{Energy level from the ground state of the relevant 
symmetric state.}  
\end{deluxetable}

\clearpage 
\setlength{\headsep}{4.5cm}
\begin{deluxetable}{cccccccccccccccc} 
\rotate 
\tablecaption{Observed Line Parameters of CH$_{\rm 3}$OH and 
C$^{\rm 34}$S \label{tbl-2}} 
\tablewidth{0pt} 
\tablehead{ \colhead{} & 
\multicolumn{11}{c}{CH$_{\rm 3}$OH\tablenotemark{a}} & \colhead{} & 
\multicolumn{3}{c}{C$^{\rm 34}$S\tablenotemark{b}} \\ 
\cline{2-12}\cline{14-16} 
\colhead{} & \multicolumn{3}{c}
{$J_{\rm K}$ = 2$_{\rm 0}$--1$_{\rm 0}$ A$^{\rm +}$} & \colhead{} & 
\multicolumn{3}{c}{$J_{\rm K}$ = 2$_{\rm -1}$--1$_{\rm -1}$ E} & \colhead{} 
& \multicolumn{3}{c}{$J_{\rm K}$ = 2$_{\rm 0}$--1$_{\rm 0}$ E} & \colhead{} 
& \multicolumn{3}{c}{$J$ = 2--1} \\ 
\cline{2-4}\cline{6-8}\cline{10-12}\cline{14-16} 
\colhead{Position} & 
\colhead{T$_{\rm A}$$^{\rm *}$} & \colhead{$\Delta$V} & 
\colhead{V$_{peak}$} & \colhead{} & \colhead{T$_{\rm A}$$^{\rm *}$} 
& \colhead{$\Delta$V} & \colhead{V$_{peak}$} & \colhead{} & 
\colhead{T$_{\rm A}$$^{\rm *}$} & \colhead{$\Delta$V} & 
\colhead{V$_{peak}$} & \colhead{} & \colhead{T$_{\rm A}$$^{\rm *}$} & 
\colhead{$\Delta$V} & \colhead{V$_{peak}$} \\ 
\colhead{} & \colhead{(K)} & \colhead{(km s$^{\rm -1}$)} & 
\colhead{(km s$^{\rm -1}$)} & \colhead{} & 
\colhead{(K)} & \colhead{(km s$^{\rm -1}$)} & 
\colhead{(km s$^{\rm -1}$)} & \colhead{} & \colhead{(K)} & 
\colhead{(km s$^{\rm -1}$)} & \colhead{(km s$^{\rm -1}$)} & \colhead{} 
& \colhead{(K)} & \colhead{(km s$^{\rm -1}$)} & 
\colhead{(km s$^{\rm -1}$)} } 
\startdata ENV\tablenotemark{c} &1.207 
&0.54 &6.70 & &0.909 &0.53 &6.63 & &0.225 &0.43 &6.65 & 
&0.446 &0.50 &6.66  \\ 
BLUE1\tablenotemark{d} 
&0.599 &0.85 &6.12 & &0.519 &1.02 &5.98 & &0.272 &0.21 &5.87 & 
&0.185 &0.36 &5.26  \\ 
BLUE2\tablenotemark{d} &0.283 &1.61 &6.50 & &0.234 &1.96 &5.48 & & - & -    
& -  & & - & - & -  \\ 
RED\tablenotemark{e} &0.740 &0.38 &6.72 & &0.544 &0.48 &6.69 & & -   & -    
& -  & &0.208 &0.24 &6.55  \\ 
RED\tablenotemark{e} &0.908 &0.37 &7.14 & &0.639 &0.32 &7.07 & & -   & -    
& -  & &0.245 &0.45 &6.95  \\ 
\enddata

\tablenotetext{a}{Rms noise levels of the CH$_{\rm 3}$OH lines are 
0.024 K, 0.059 K, 0.041 K, and 0.052 K in ENV, BLUE1, BLUE2, and 
RED, respectively.}
\tablenotetext{b}{Rms noise levels of the C$^{\rm 34}$S line are 
0.026 K, 0.058 K, 0.055 K, and 0.062 K in ENV, BLUE1, BLUE2, and 
RED, respectively.}
\tablenotetext{c}{Line parameters are determined by a 
single Gaussian fit to the spectra.} 
\tablenotetext{d}{Line parameters 
are determined by measuring the peak intensity, peak velocity, and the 
FWHM of the spectra.} 
\tablenotetext{e}{Line parameters are 
determined by two-components Gaussian fit to the spectra.}

\end{deluxetable}

\clearpage 
\setlength{\headsep}{4.5cm}

\begin{deluxetable}{cccccccccccc} 
\rotate 
\tablecaption{Physical and Chemical Properties of the CH$_{\rm 3}$OH 
Components \label{tbl-3}} 
\tablewidth{0pt} 
\tablehead{
\colhead{Component} & \colhead{$N_{\rm H_{2}}$\tablenotemark{a}} & 
\colhead{$\tau$\tablenotemark{b,c}} 
& \colhead{$N_{\rm CH_{3}OH}$\tablenotemark{c,d}} 
& \colhead{X$_{\rm CH_{3}OH}$\tablenotemark{c,d}} 
& \colhead{V$_{\rm CH_{3}OH}$\tablenotemark{e}} 
& \colhead{$\theta$$_{\rm FWHM}$\tablenotemark{f}} 
& \colhead{$D$\tablenotemark{f}} & 
\colhead{$\Delta$V\tablenotemark{e}} & 
\colhead{M$_{vir}$\tablenotemark{g}} & \colhead{M$_{\rm 
LTE}$\tablenotemark{g}} & \colhead{Mom.\tablenotemark{g,h}} \\ 
\colhead{} & \colhead{($\times$ 10$^{\rm 22}$ cm$^{\rm -2}$)} & 
\colhead{} & \colhead{($\times$ 10$^{\rm 14}$ cm$^{\rm -2}$)} & 
\colhead{($\times$ 10$^{\rm -8}$)} & \colhead{(km s$^{\rm -1}$)} & 
\colhead{($\arcsec$ $\times$ $\arcsec$)} & \colhead{(pc)} & 
\colhead{(km s$^{\rm -1}$)} & \colhead{(M$_{\odot}$)} & 
\colhead{(M$_{\odot}$)} & \colhead{(M$_{\odot}$ km s$^{\rm -1}$)} } 
\startdata ENV  &4.26  &0.43  &0.89 &0.2 &6.7  &106 $\times$ 51 
&0.050  &0.53 &2.4  &2.300   & -   \\
BLUE1    &0.54  &0.19  &1.16  &2.1  
&5.9  &27 $\times$ 21  &0.016  &0.84 &1.5  &0.064 &0.058 \\ 
BLUE2  &0.57  &0.09 &0.62 &1.1  &6.0  &74 $\times$ 26  &0.030  &1.61 &8.6  
&0.085 &0.073 \\ 
RED       &0.73  &0.31  &0.83 &1.1  &7.1  &56 $\times$ 19  
&0.022  &0.35 &0.7 &0.069 &0.040 \\ \enddata

\tablenotetext{a}{Molecular hydrogen column densities are estimated 
from the 1.3 mm dust continuum map by Motte \& Andr\'e (2001).}

\tablenotetext{b}{Peak optical depth of the 
2$_{\rm 0}$--1$_{\rm 0}$ A$^{\rm +}$ line.}

\tablenotetext{c}{
On the assumption of the excitation temperature of 
10 K.}

\tablenotetext{d}{On the assumption of equal abundance of A- and 
E-symmetric species of CH$_{\rm 3}$OH.}

\tablenotetext{e}{We picked up the CH$_{\rm 3}$OH spectra toward the 
central positions of each identified component, and made Gaussian 
fittings to the spectra to measure the velocity and the line width 
($\Delta$V) of each gas component. If the line profiles are not 
well-fitted by a Gaussian, we simply measure the FWHM line width. In 
estimating the line widths, we deconvolved the measured line width 
with the velocity resolution.} 

\tablenotetext{f}{We measured the size 
of each gas components by 2-dimensional Gaussian fittings to the 
integrated intensity maps in Figure 3. The deconvolution with the beam 
size were performed to obtain the size. Here, the size ($D$) of the 
condensations is the geometrical mean of the major and minor axis 
($\theta$$_{\rm FWHM}$) from the 2-dimensional Gaussian fit.} 

\tablenotetext{g}{See text.}

\tablenotetext{h}{Corrections for the inclination of the outflow to the 
line of sight ($\sim$ 40$^{\circ}$ : Andr\'e et al. 1999a) are applied.}

\end{deluxetable}

\clearpage 
\begin{deluxetable}{lccc} 
\tablecaption{Offset Velocities of the 
CH$_{\rm 3}$OH Condensations\label{tbl-4}} 
\tablewidth{0pt} 
\tablehead{
\colhead{Condensations} & \colhead{R} & 
\colhead{V$_{off}$} & \colhead{V$_{g}$} \\ 
\colhead{} & \colhead{(pc)} & \colhead{(km s$^{\rm - 1}$)} & 
\colhead{(km s$^{\rm -1}$)} }
 
\startdata BLUE1 &0.055 &1.03  &0.60  \\ 
           BLUE2 &0.027 &0.87  &0.86  \\ 
           RED   &0.041 &0.45\tablenotemark{a} &0.71\\ 
\enddata

\tablenotetext{a}{No correction for the inclination angle is applied.}

\end{deluxetable}

\clearpage 
\begin{deluxetable}{lccccccccccc} \tabletypesize{\scriptsize} 
\tablecaption{Comparison of Outflow Sources with CH$_{\rm 3}$OH
Observations \label{tbl-5}} 
\tablewidth{0pt} \tablehead{ \colhead{} & 
\multicolumn{2}{c}{Protostar} & \colhead{} & 
\multicolumn{3}{c}{CO Outflow} & \colhead{} & \multicolumn{3}{c}{CH$_{\rm 3}$OH} & 
\colhead{} \\ \cline{2-3}\cline{5-7}\cline{9-11} \colhead{Source} & 
\colhead{L$_{bol}$} & \colhead{T$_{bol}$} & \colhead{} & 
\colhead{V$_{\rm flow}$\tablenotemark{a,b}} & 
\colhead{Momentum\tablenotemark{b}} & \colhead{T$_{dyn}$} & 
\colhead{} & \colhead{X\tablenotemark{c}} & 
\colhead{2$_{\rm 0}$--1$_{\rm 0}$ A$^{\rm +}$ $\Delta$V} & 
\colhead{V$_{off}$\tablenotemark{b}} & 
\colhead{Ref.\tablenotemark{d}} \\ 
\colhead{} & 
\colhead{(L$_{\odot}$)} & \colhead{(K)} & \colhead{} 
& \colhead{(km s$^{\rm -1}$)} & \colhead{(M$_{\odot}$ km s$^{\rm -1}$)} & 
\colhead{(yr)} & \colhead{} & \colhead{} & \colhead{(km s$^{\rm -1}$)} 
& \colhead{(km s$^{\rm -1}$)} & \colhead{} }

\startdata IRAM 04191     &0.15 &18 & &5.5 &0.12 &8000  &  &$<$10 
&0.85  &1.0  &1,2,3 \\ L1157          &5.8  &42 & &49  &30   &15000 &  
&400   &5.9   &36   &4,5,6 \\ NGC1333 IRAS 2 &7.2  &26 & &22  &0.64 
&12000 &  &300   &7.2   &18   &7,8,9 \\ 
BHR 71         &9    &35 & &28  &11   &10000 &  &40    &4.1   
&31   &10,11 \\ \enddata

\tablenotetext{a}{The flow velocity (V$_{\rm flow}$) is determined by 
dividing the momentum by the outflow mass, that is, the flow velocity 
weighted by the mass.}

\tablenotetext{b}{Corrections for the effect of the inclination of the 
outflow are applied.}

\tablenotetext{c}{CH$_{\rm 3}$OH abundance enhancement toward the 
outflow as compared to that in the protostellar position.}

\tablenotetext{d}{$References:$ (1) This work, (2) Andr\'e et al. 1999a, 
(3) Ohashi 1991, (4) Bachiller et al. 2001, (5) Shirley et al. 2000, (6) 
Bachiller et al. 1995, (7) Knee \& Sandell 2000, (8) Bachiller et al. 
1998, (9) Sandell et al. 1994, (10) Garay et al. 1998, 
(11) Bourke et al. 1997}

\end{deluxetable}


\clearpage


\begin{thebibliography}{} 

\bibitem[Adams et al.\ 1987]{ada87} Adams, F. C., Lada, C. J., 
\& Shu, F. H. 1987, \apj, 312, 788 
\bibitem[Adams 1990]{ada90} Adams, F. C.  1990, \apj, 363, 578 
\bibitem[Aikawa et al.\ 2001]{aik01} Aikawa, Y., 
Ohashi, N., Inutsuka, S., Herbst, E., \& Takakuwa, S. 2001, \apj, 552, 639 
\bibitem[Andr\'e et al.\ 1999a]{an99a} Andr\'e, P., Motte, F., \& 
Bacmann, A. 1999a, \apjl, 513, L57 
\bibitem[Andr\'e et al.\ 1999b]{an99b} Andr\'e, P., Motte, F., Bacmann, A., 
\& Belloche, A. 1999b, 
in Proceedings of the International Conference ``Star Formation 1999'', 
Edited by T. Nakamoto, Published by The Nobeyama Radio Observatory, 145
\bibitem[Avery \& Chiao 1996]{ave96} Avery, L. W. \& Chiao, M. 
1996, \apj, 463, 642
\bibitem[Bachiller et al.\ 1991]{bac91} Bachiller, 
R., Martin-Pintado, J., \& Fuente, A. 1991, \aap, 243, L21 
\bibitem[Bachiller et al.\ 1995]{bac95} Bachiller, R., Liechti, S., 
Walmsley, C. M., \& Colomer, F. 1995, \aap, 295, L51 
\bibitem[Bachiller \& Guti\'errez 1997]{bac97} Bachiller, R., \&
Guti\'errez, M. P. 1997, \apjl, 487, L93
\bibitem[Bachiller et al.\ 1998]{bac98} Bachiller, R., Codella, C., 
Colomer, F., Liechti, S., \& Walmsley, C. M. 1998, \aap, 335, 266 
\bibitem[Bachiller et al.\ 2001]{bac01} Bachiller, R., Guti\'errez, M. 
P., Kumar, M. S. N., \& Tafalla, M. 2001, \aap, 372, 899
\bibitem[Belloche et al.\ 2002]{bel02} Belloche, A., Andr\'e, P., 
Despois, D., \& Blinder, S. 2002, \aap, 393, 927
\bibitem[Benson \& Myers 1989]{ben89} Benson, P. J., \& Myers, P. C. 
1989, \apjs, 71, 89
\bibitem[Bergin et al.\ 1998]{ber98} Bergin, E. A., Melnick, G. J., \&
Neufeld, D. A. 1998, \apj, 499, 777
\bibitem[Blake et al.\ 1994]{bla94} 
Blake, G. A., Van Dishoeck, E. F., Jansen, D. J., 
Groesbeck, T. D., \& Mundy, L. G. 1994, \apj, 428, 680 
\bibitem[Blake et al.\ 1995]{bla95} Blake, G. 
A., Sandell, G., Van Dishoeck, E. F., Groesbeck, T. D., 
Mundy, L. G., \& Aspin, C. 1995, \apj, 441, 689 
\bibitem[Bourke et al.\ 1997]{bou97} Bourke, T. L., Garay, G., Lehtinen, 
K. K., K\"ohnenkamp, I., Launhardt, R., Nyman, L.-\AA., May, J., Robinson, 
G., \& Hyland, A. R. 1997, \apj, 476, 781
\bibitem[Buckle \& Fuller 2002]{buc02} Buckle, J. V., \& Fuller, G. A. 
2002, \aap, 381, 77
\bibitem[Ceccarelli et al.\ 2000]{cec00} Ceccarelli, C., Castets, A., 
Caux, E., Hollenbach, D., Loinard, L., Molinari, S., Tielens, A.G.G.M. 2000, 
\aap, 355, 1129 
\bibitem[Dutrey et al.\ 1997]{dut97} Dutrey, A., Guilloteau, S., \& 
Bachiller, R. 1997, \aap, 325, 758 
\bibitem[Elias 1978]{eli78} Elias, J. H. 1978, \apj, 224, 857 
\bibitem[Friberg et al.\ 1988]{fri88} Friberg, P., Madden, S. C., 
Hjalmarson, A., \& Irvine, W. M. 1988, \aap, 195, 281 
\bibitem[Garay et al.\ 1998]{gar98} Garay, G., K\"ohnenkamp, I., 
Bourke, T. L., Rodr\'{\i}guez, L. F., \& Lehtinen, K. K. 1998,
\apj, 509, 768 
\bibitem[Gibb \& Davis 1998]{gib98} Gibb, A. G., \& Davis, C. J. 1998, 
\mnras, 298, 644
\bibitem[Goldreich \& Kwan 1974]{gol74} Goldreich, P., \& Kwan, J. 1974, 
\apj, 189, 441
\bibitem[Goodman et al.\ 1993]{goo93} Goodman, A. A., Benson, P. J., 
Fuller, G. A., \& Myers, P. C. 1993, \apj, 406, 528 
\bibitem[Gueth et al.\ 1998]{gue98} Gueth, F., Guilloteau, S., \&
Bachiller, R. 1998, \aap, 333, 287
\bibitem[Hatchell et al.\ 1999a]{ha99a} Hatchell, J., Fuller, G. A., 
\& Ladd, E. F. 1999a, \aap, 344, 687
\bibitem[Hatchell et al.\ 1999b]{ha99b} Hatchell, J., Fuller, G. A., 
\& Ladd, E. F. 1999b, \aap, 346, 278 
\bibitem[Hirano et al.\ 2001]{hir01} Hirano, N., Mikami, H., 
Umemoto, T., Yamamoto, S., \& Taniguchi, Y. 2001, \apj, 547, 899
\bibitem[Hollenbach 1997]{hol97} Hollenbach, D. 1997, in IAU Symp. 
182, Herbig-Haro Flows and the Birth of Low Mass Stars, ed. B. Reipurth 
\& C. Bertout (Dordrecht:Kluwer), 181 
\bibitem[Kaufman \& Neufeld 1996a]{ka96a} Kaufman, M. J., \& Neufeld, D. A. 
1996a, \apj, 456, 250 
\bibitem[Kaufman \& Neufeld 1996b]{ka96b} Kaufman, M. J., \& 
Neufeld, D. A. 1996b, \apj, 456, 611 
\bibitem[Knee \& Sandell 2000]{kne00} Knee, L. B. G., \& Sandell, G. 2000, 
\aap, 361, 671 
\bibitem[Kuiper et al.\ 1996]{kui96} Kuiper, T. B. H., Langer, W. D., \& 
Velusamy, T. 1996, \apj, 468, 761 
\bibitem[Lada 1991]{lad91} Lada, C. J. 1991, in The Physics of 
Star Formation and Early Stellar Evolution, ed. C. J. Lada \& N. D. 
Kylafis, Dordrecht: Reidel, 1991, 329 
\bibitem[Langer et al.\ 1995]{lan95} Langer, W. D., Velusamy, T., 
Kuiper, T. B. H., Levin, S., Olsen, E., \& Migenes, V. 1995, \apj, 453, 293 
\bibitem[Lee et al.\ 2002]{lee02} Lee, C.-F., Mundy, L. G., 
Stone, J. M., \& Ostriker, E. C. 2002, \apj, 576, 294
\bibitem[Mikami et al.\ 1992]{mik92} Mikami, H., Umemoto, T., 
Yamamoto, S., \& Saito, S.  1992, \apjl, 392, L87 
\bibitem[Mizuno et al.\ 1994]{miz94} Mizuno, A., 
Onishi, T., Hayashi, M., Ohashi, N., Sunada, K., Hasegawa, T., \& Fukui, Y. 
1994, \nat, 368, 719 
\bibitem[Momose et al.\ 1996]{mom96} Momose, M., Ohashi, N., Kawabe, R., 
Hayashi, M., \& Nakano, T. 1996, \apj, 470, 1001 
\bibitem[Motte \& Andr\'e 2001]{mot01} Motte, 
F., \& Andr\'e, P. 2001, \aap, 365, 440 
\bibitem[Myers et al.\ 1995]{mye95} Myers, P. C., 
Bachiller, R., Caselli, P., Fuller, G. A., Mardones, D., Tafalla, M., \& 
Wilner, D. J. 1995, \apjl, 449, L65 
\bibitem[Nakano et al.\ 1995]{nak95} 
Nakano, T., Hasegawa, T., \& Norman, C. 1995, \apj, 450, 183 
\bibitem[Ohashi 1991]{oha91} 
Ohashi, N. 1991, PhD thesis, Nagoya University 
\bibitem[Ohashi et al.\ 1997a]{oh97a} Ohashi, N., Hayashi, M., 
Ho, P. T. P., \& Momose, M. 1997a, \apj, 475, 211 
\bibitem[Ohashi et al.\ 1997b]{oh97b} Ohashi, N., 
Hayashi, M., Ho, P. T. P., Momose, M., Tamura, M., Hirano, N., \& Sargent, 
A. I. 1997b, \apj, 488, 317 
\bibitem[Ohashi et al.\ 1999]{oha99} Ohashi, 
N., Lee, S. W., Wilner, D. J., \& Hayashi, M. 1999, \apjl, 518, L41 
\bibitem[Ohashi 2000]{oha00} Ohashi, N. 2000, in Astrochemistry: 
>From Molecular Clouds to Planetary Systems, ed. Y. C. Minh \& E. F. van 
Dishoeck (San Francisco: ASP), 61 
\bibitem[Peng et al.\ 1998]{pen98} Peng, R., Langer, W. D., Velusamy, T., 
Kuiper, T. B. H., \& Levin, S. 1998, \apj, 497, 842 
\bibitem[Pineau des For\^ets et al.\ 1997]{pin97} Pineau des For\^ets, G., 
Flower, D. R., \& Chi\`eze, J. -P. 1997, in IAU Symp. 182, 
Herbig-Haro Flows and the Birth 
of Low Mass Stars, ed. B. Reipurth \& C. Bertout (Dordrecht:Kluwer), 199 
\bibitem[Pratap et al.\ 1997]{pra97} Pratap, P., Dickens, J. E., Snell, R. 
L., Miralles, M. P., Bergin, E. A., Irvine, W. M., \& Schloerb, F. P. 1997, 
\apj, 486, 862 
\bibitem[Raga \& Cabrit 1993]{rag93} Raga, A., \& 
Cabrit, S. 1993, \aap, 278, 267 
\mnras, 306, 691 
\bibitem[Saito et al.\ 1999]{sai99} Saito, M., Sunada, 
K., Kawabe, R., Kitamura, Y., \& Hirano, N. 1999, \apj, 518, 334 
\bibitem[Sandell et al.\ 1994]{san94} Sandell, G., Knee, L. B. G., Aspin, 
C., Robson, I. E., \& Russell, A. P. G. 1994, \aap, 285, L1 
\bibitem[Sandford \& Allamandola 1993]{san93} Sandford, S. A., \& 
Allamandola, L. J. 1993, \apj, 417, 815 
\bibitem[Schilke et al.\ 1997]{sch97} Schilke, P., Walmsley, C. M., 
Pineau des For\^ets, G., \& Flower, D. R. 1997, \aap, 321, 293 
\bibitem[Scoville \& Solomon 1974]{sco74} Scoville, N. Z., \& 
Solomon, P. M. 1974, \apjl, 187, L67 
\bibitem[Shirley et al.\ 2000]{shi00} Shirley, Y. L., Evans II, N. J., 
Rawlings, J. M. C., \& Gregersen, E. M. 2000, \apjs, 131, 249 
\bibitem[Swade 1989]{swa89} Swade, D. A. 1989, \apj, 345, 828 
\bibitem[Takakuwa et al. \ 1998]{tak98} Takakuwa, S., Mikami, H., 
\& Saito, M. 1998, \apj, 501, 723 
\bibitem[Takakuwa et al.\ 2000]{tak00} Takakuwa, S., Mikami, 
H., Saito, M., \& Hirano, N. 2000, \apj, 542, 367 
\bibitem[Umemoto et al.\ 1992]{ume92} Umemoto, T., Iwata, T., Fukui, Y., 
Mikami, H., Yamamoto, S., Kameya, O., \& Hirano, N. 1992, \apjl, 392, L83 
\bibitem[Umemoto et al.\ 1999]{ume99} Umemoto, T., Mikami, H., 
Yamamoto, S., \& Hirano, N. 1999, \apjl, 525, L105 
\bibitem[Velusamy \& Langer 1998]{vel98} Velusamy, T., \& Langer, W. D. 
1998, \nat, 392, 685 
\bibitem[Xu \& Lovas 1997]{xu97} Xu, L., \& Lovas, F. J. 1997, J. Phys. Chem. Ref. Data, 
26, 17
\bibitem[Zhou 1992]{zho92} Zhou, S. 1992, \apj, 394, 204 
\bibitem[Zhou et al.\ 1994]{zho94} Zhou, S., Evans II, N. J., Wang, Y., 
Peng, R., \& Lo, K. Y., 1994, \apj, 433, 131 
\end{thebibliography}
\end{document}